\documentclass[5p,times,merge,number,sort&compress]{elsarticle}

\usepackage[T1]{fontenc}
\usepackage[utf8]{inputenc}
\usepackage[english]{babel}

\usepackage{graphicx}
\usepackage{isotope}
\usepackage{physics}
\usepackage{xcolor}
\usepackage{subcaption}
\usepackage{hyperref}
\usepackage{amsmath}
\usepackage{multirow}

\DeclareMathAlphabet{\curly}{OMS}{cmsy}{m}{n}

\newcommand*{\elem}[2]{\ensuremath{\isotope[#2]{\mathrm{#1}}}}

\definecolor{FGOrange}{rgb}{1.0, 0.5490196078431373, 0.0}
\definecolor{FGgreen}{rgb}{.0, 0.5, 0.0}
\definecolor{FGred}{rgb}{1.0, 0.0, 0.0}

\definecolor{FGredgray1}{rgb}{.5, 0.5, 0.5}
\definecolor{FGredgray2}{rgb}{.6016, 0.3359, 0.3359}
\definecolor{FGredgray3}{rgb}{.7031, 0.1680, 0.1680}
\definecolor{FGredgray4}{rgb}{.6797, 0.0742, 0.0742}
\definecolor{FGredgray5}{rgb}{.5586, 0.0352, 0.0352}
\definecolor{FGredgray6}{rgb}{.4492, 0., 0.0}

\begin{document}

\title{Machine Learning for the Prediction of Converged Energies\\from Ab Initio Nuclear Structure Calculations}

\address[tud]{Institut f\"ur Kernphysik, Fachbereich Physik, Technische Universit\"at Darmstadt, Schlossgartenstr. 2, 64289 Darmstadt, Germany}
\address[hfhf]{Helmholtz Forschungsakademie Hessen f\"ur FAIR, GSI Helmholtzzentrum, 64289 Darmstadt, Germany}

\author[tud]{Marco~Kn\"oll}
\ead{mknoell@theorie.ikp.physik.tu-darmstadt.de}
\author[tud]{Tobias~Wolfgruber}
\author[tud]{Marc~L.~Agel}
\author[tud]{Cedric~Wenz}
\author[tud,hfhf]{Robert~Roth}
\ead{robert.roth@physik.tu-darmstadt.de}

\date{\today}

\begin{abstract}

\noindent The prediction of nuclear observables beyond the finite model spaces that are accessible through modern \textit{ab initio} methods, such as the no-core shell model, pose a challenging task in nuclear structure theory.
It requires reliable tools for the extrapolation of observables to infinite many-body Hilbert spaces along with reliable uncertainty estimates.
In this work we present a universal machine learning tool capable of capturing observable-specific convergence patterns independent of nucleus and interaction.
We show that, once trained on few-body systems, artificial neural networks can produce accurate predictions for a broad range of light nuclei.
In particular, we discuss neural-network predictions of ground-state energies from no-core shell model calculations for \elem{Li}{6}, \elem{C}{12} and \elem{O}{16} based on training data for \elem{H}{2}, \elem{H}{3} and \elem{He}{4} and compare them to classical extrapolations.
\end{abstract}

\maketitle

\paragraph{Introduction}

The major goal of nuclear structure theory is the accurate description of nuclear properties based on the underlying strong interaction. The low-energy properties of these complicated quantum many-body systems emerge from the interactions among nucleons as the relevant degrees of freedom. 
While the construction of these interactions poses a challenge in itself, also the solution of the nuclear many-body problem with controlled uncertainties is a formidable conceptual and computational task. Several successful \textit{ab initio} methods are available nowadays, such as the no-core shell model (NCSM) \cite{BarNa13,NaQu09,Roth09,ZheBa93}, the coupled-cluster (CC) method \cite{KoDe04}, the self-consistent Green's function (SCGF) approach \cite{DiBa04}, the in-medium similarity renormalization group (IM-SRG) \cite{HeBo16}, or lattice and continuum quantum Monte Carlo (QMC) methods \cite{CaGa15}.
Besides the latter, all of these methods are built on the expansion of the many-body problem in a discrete many-body basis that is constructed from an underlying single-particle basis. To render the computational problem finite, various basis truncations are employed that define finite-dimensional model spaces.
For \textit{ab initio} methods the truncations are constructed in such a way that calculations for increasing model-space dimension eventually converge to the exact result for the full many-body Hilbert space.

In the past two decades \textit{ab initio} methods have proven to provide accurate results for a range of observables and nuclei based on nucleon-nucleon (NN) and three-nucleon (3N) interactions from chiral effective field theory (EFT) \cite{MaEn11,EpHa09,HuVo20,LENPIC21,HeBo13}.
Besides their success, all of these methods suffer from a rapid growth of the underlying basis and the resulting model spaces. Particularly for configuration interaction approaches, like the NCSM, the factorial growth of the model-space dimension with increasing particle number \cite{MaVa09} sets severe limits for converged calculations. Even with access to high performance computing and methods to accelerate convergence, such as the similarity renormalization group (SRG) \cite{BoFu07,BoFu10},  one is inevitably confronted with incomplete convergence of the many-body calculation. Hence, there is a need for extrapolation procedures that provide robust predictions for the converged observables along with reliable uncertainty estimates.

Traditional extrapolation schemes typically rely on empirical exponential or polynomial parametrizations of the model-space dependence of observables \cite{Roth09,MaVa09,BoFu07b}. Recent physics-motivated parametrizations, like the infrared extrapolation schemes derived from effective theories \cite{FuHa12,CoAv12,MoEk13,FuMo14,WeFo15}, have proven successful in specific cases, but impose additional constraints on the many-body calculations. 

With increasing popularity of machine learning, artificial neural networks (ANNs) have entered the field of nuclear structure physics, e.g., through large-scale approaches based on experimental data \cite{AtMa04,AkBa13}, theoretical applications in the form of Bayesian machine-learning \cite{UtCh16}, neural-network quantum states \cite{AdCo21,EkFo19,Yosh20}, and many more.
For a comprehensive overview and further reading we refer to \cite{Clar99,BoAm22}.
While ANNs have excelled in classification and interpolation tasks, precise extrapolations remain challenging \cite{XuZh20}.
However, first applications to NCSM and CC calculations have demonstrated the potential of machine learning as an extrapolation tool supplementing \textit{ab initio} many-body methods \cite{NeVa19,JiHa19}.
So far, these applications extrapolate NCSM or CC ground-state observables by emulating their model-space dependence, more precise, their functional dependence on the model-space truncation parameter and the harmonic-oscillator (HO) frequency $\hbar\Omega$ of the underlying single-particle basis. The ANNs are trained to mimic this functional dependency for a specific interaction, nucleus, and eigenstate and are then used to extrapolate the observable to a sufficiently large model space. These approaches essentially replace the traditional exponential parametrization of the many-body data by an ANN, which contains many more parameters. The training process for the ANN requires a sufficient amount of data, i.e., a large number of many-body calculations for a specific target nucleus, interaction, and eigenstate with different model spaces. This can easily become a bottleneck for the applicability of this scheme.
While the problem can partially be addressed through suitable interpolations of the training data as shown in \cite{JiHa19}, no data from large model spaces or actually converged results will ever inform the network. Moreover, the resulting ANN, by construction, is only useful for an extrapolation for the specific nucleus, interaction, and state it was trained for. 

\begin{figure*}
    \centering
     \begin{subfigure}{0.49\textwidth}
         \centering
         \includegraphics[width=\textwidth]{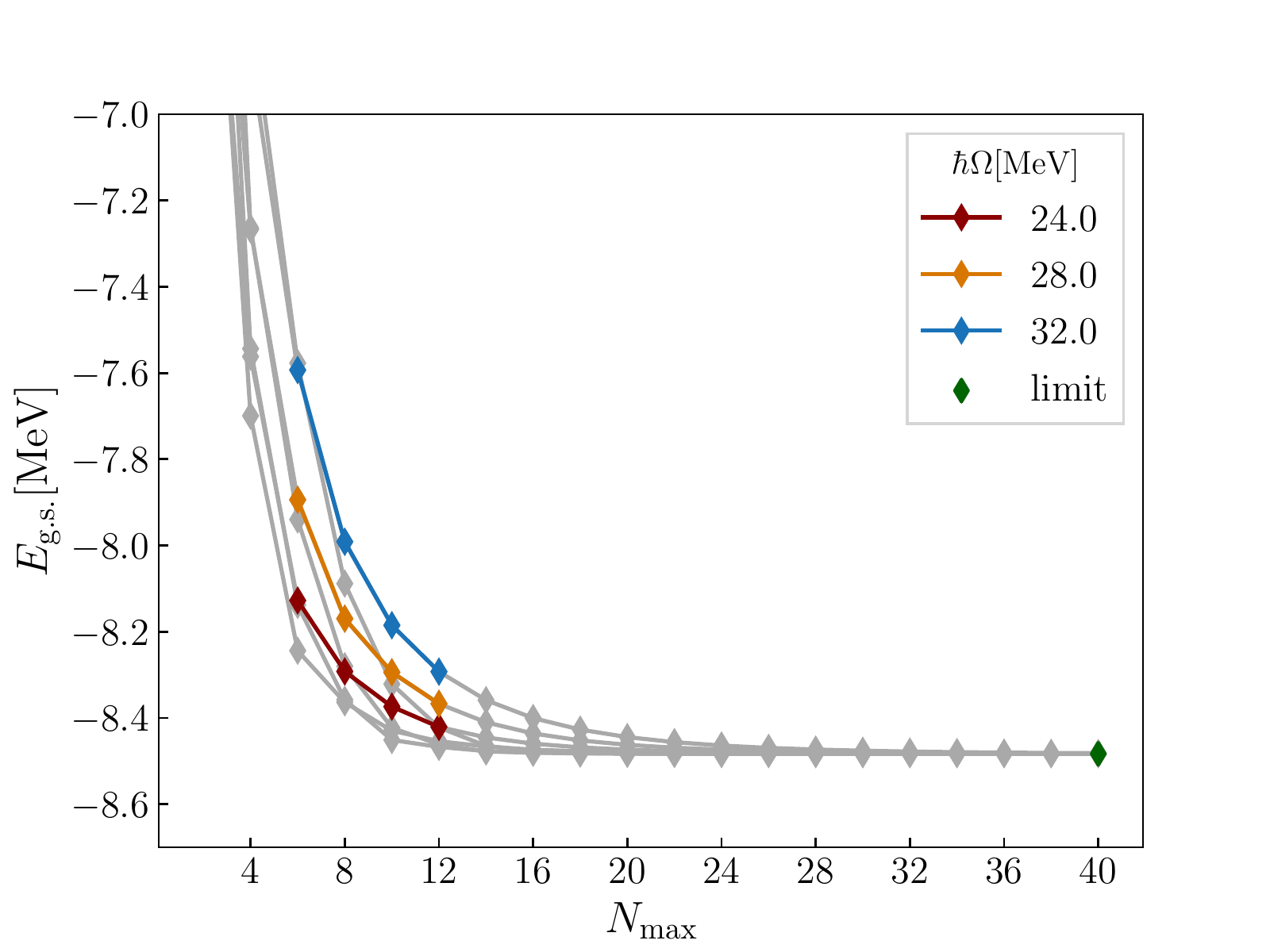}
         \caption{NCSM results}
         \label{fig:NCSM}
     \end{subfigure}
     \begin{subfigure}{0.49\textwidth}
         \centering
         \includegraphics[width=\textwidth]{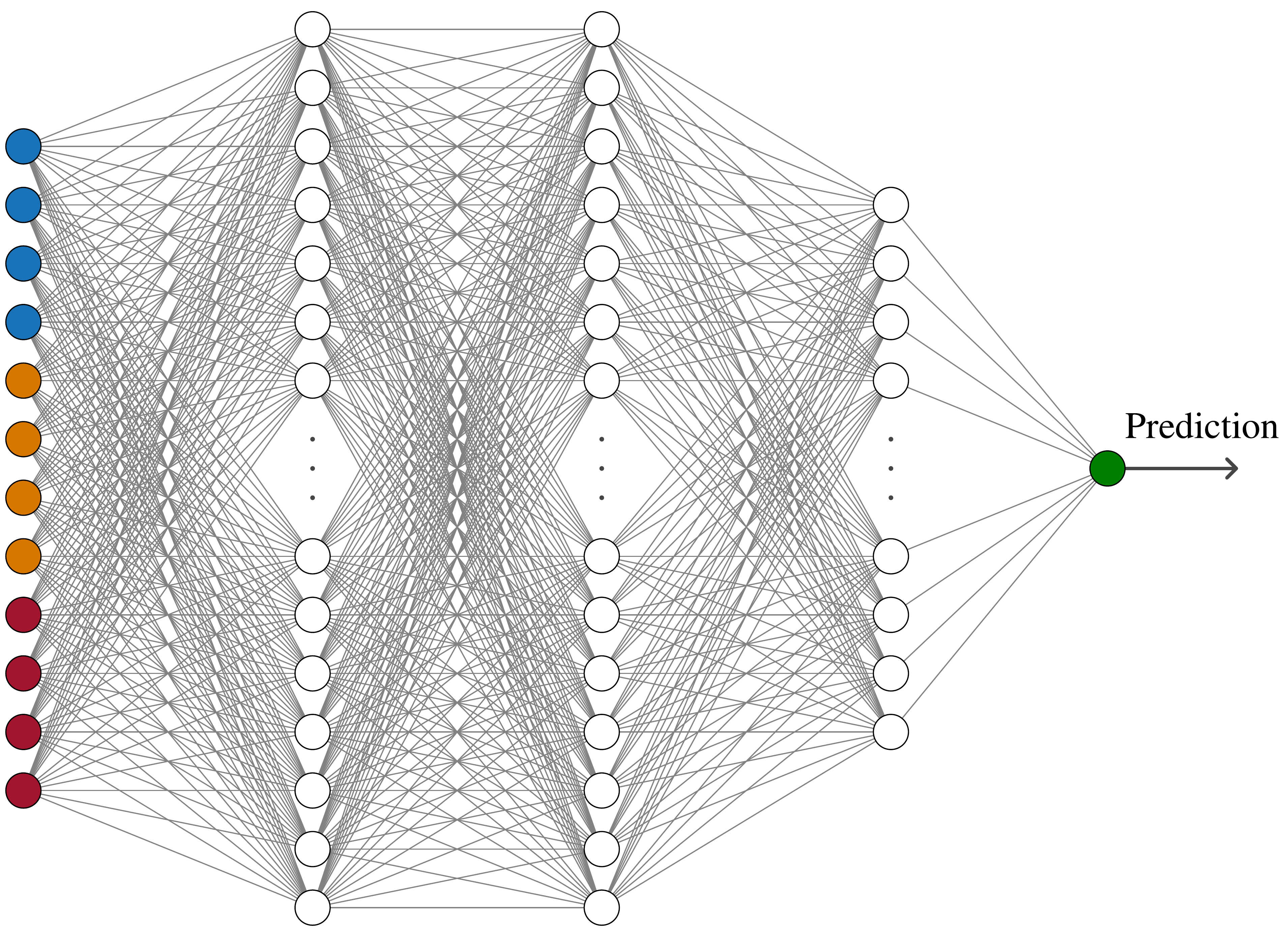}
         \caption{network topology}
         \label{fig:ANN}
     \end{subfigure}
     \caption{(a) Exemplary results from NCSM calculations for the ground-state energy of \elem{H}{3}. Shown in gray is the training data for one given interaction. The colored set of values for four consecutive $N_\mathrm{max}$ and three different HO frequencies together with the limit represents a single training sample for the ANNs. (b) Visualization of the topology of the fully connected feedforward ANN employed in this work. The colors in the first and last layer correspond to the colored sample in (a).}
\end{figure*}

We present a new approach, build on a universal ANN for the prediction of the converged values of a specific observable for arbitrary nuclei, interactions, and eigenstates from sequences of NCSM calculations. The basic idea is akin to a pattern recognition task: Short sequences of NCSM results with increasing basis truncation parameter for different HO frequencies $\hbar\Omega$ form a pattern that allows an experienced many-body practitioner to guess what the converged value will be. For this, one does not even have to know which nucleus, interaction, or eigenstate the data was generated for---it is just the pattern of converging sequences that holds the relevant information. To extract quantitative predictions, we design an ANN that uses the raw many-body results for short $N_{\max}$ sequences and several frequencies $\hbar\Omega$ as input and outputs the converged value of the observable. We train the ANN with calculations for few-body systems $(A\leq 4)$, which can be fully converged in the NCSM, with a large number of different Hamiltonians. 
Once trained on a huge set of training data, the ANNs serve as universal tool for the prediction of the converged observable for expensive many-body runs, where only a very limited amount of data is available. In this work we discuss predictions of ground-state energies from ANNs that have been trained on converged NCSM calculations for the few-body systems \elem{H}{2}, \elem{H}{3} and \elem{He}{4}. We subsequently apply the ANNs also to heavier p-shell nuclei, i.e., \elem{Li}{6}, \elem{C}{12} and \elem{O}{16}.

\paragraph{No-Core Shell Model}

Our many-body method of choice is the NCSM, in which the stationary Schrödinger equation is cast into a matrix eigenvalue problem
\begin{align}
    \sum_j\mel*{\phi_i}{H}{\phi_j}\braket*{\phi_j}{\psi_n} = E_n\braket{\phi_i}{\psi_n} \quad \forall i,
\end{align}
with Hamiltonian $H$, its energy eigenvalues $E_n$, and the corresponding eigenstates $\ket{\psi_n}$.
The eigenstates are expanded in a set of Slater determinants $\{\ket{\phi_i}\}$, which are constructed from HO single-particle states, introducing a dependence on the HO frequency $\hbar\Omega$.
Furthermore, the model space is truncated by introducing an upper limit to the number of harmonic-oscillator excitation quanta $N_\mathrm{max}$. The convergence behavior is controlled by the $N_\mathrm{max}$ truncation parameter and the HO frequency $\hbar\Omega$.

The Hamiltonian consists of a kinetic energy term and NN and 3N interactions obtained from chiral EFT, where the NN interaction comes with a short-range repulsion which induces strong short-range correlations.
In a Slater determinant basis, the representation of these short-range correlations requires large model spaces, as they connect high and low-momentum states leading to slow many-body convergence. 
The convergence can be accelerated by using an SRG transformation of the Hamiltonian \cite{RoNe10,FuHe13,RoCa14}, which decouples high and low-momentum states and, thus, prediagonalizes the Hamiltonian. Therefore, with increasing flow parameter $\alpha$ of the SRG evolution, the model-space convergence is accelerated.
In this work, we utilize the SRG transformation to generate a larger variety of convergence patterns for the training of the ANNs in order to make them robust. 

For nuclei with $A\leq 4$ we employ the Jacobi-NCSM~\cite{LiMe16}, a reformulation of the NCSM in a HO basis constructed in Jacobi coordinates. It enables the efficient computation of few-nucleon systems up to very large values of $N_\mathrm{max}$, reaching fully converged ground-state energies. 
An example for a set of Jacobi-NCSM calculations for \elem{H}{3} is presented in Fig.~\ref{fig:NCSM}, which shows sequences of ground-state energies as a function of $N_\mathrm{max}$ for multiple HO frequencies that converge towards the same limit.
Due to the variational character of the NCSM approach, the ground-state energies always exhibit a monotonously decreasing convergence. 
A reasonable approximation for the $N_{\max}$-dependence of the energies for a single HO frequency is an exponential function.
Therefore, predictions of the converged energy are traditionally extracted through exponential fits to manually selected frequencies \cite{MaVa09,JuMa13,LENPIC21}, where the frequency with the minimum energy at the largest $N_\mathrm{max}$ accessible yields the nominal extrapolated result. The extrapolations for neighboring frequencies serve as uncertainty estimates. 
Clearly, this is an empirical and potentially biased process that might strongly depend on the chosen frequencies.
Given the simple and systematic convergence patterns, ANNs seem to provide a powerful alternative to these schemes.

\paragraph{Artificial Neural Networks}

Generally speaking, ANNs are useful tools for pattern recognition.
Being loosely modeled after the human brain, they consist of processing units called neurons arranged in layers~\cite{goodfellow2016deep}.
In a dense feedforward network, as employed in this work, all neurons in a given layer receive inputs from all neurons of the preceding layer and cast their output to all neurons of the following layer.
A schematic representation of the network topology is shown in Fig.~\ref{fig:ANN}.
The output $x^\mathrm{out}_i$ of the $i$-th neuron in a specific layer is calculated from the inputs $x^\mathrm{in}_j$ via
\begin{align}
    x^\mathrm{out}_i=\sigma\qty(\sum_j x^\mathrm{in}_j w_{ji} + b_i)
\end{align}
with an activation function $\sigma$ along with weights $w_{ji}$ and biases $b_i$ which correspond to the strengths of the connections between the neurons and their internal thresholds, respectively.
These parameters are usually initialized randomly and are adjusted iteratively during the training process. 
The training or supervised learning can be understood as a high-dimensional fit to the training data based on a minimization of the loss function, which provides a measure for the deviation between the network outputs and the target values.

In each iteration a batch of training samples is passed through the network.
This step is known as forward pass.
Next, the loss is determined from the network's predictions for all samples in the batch.
During the following backward pass, the weights and biases of the ANN are adjusted through a back-propagation algorithm~\cite{Rojas96}.
This key process allows for a systematic optimization of every single parameter based on the gradient of the loss function, similar to steepest descend, in order to minimize the loss.
Additionally, the adjustments are multiplied by a factor called learning rate, which controls the step size of the optimization process.
Using a learning rate scheduler this factor will be reduced if the loss plateaus to achieve higher accuracy.
Once trained, the network should have captured the convergence patterns and should provide an accurate description of the training data, leading to good predictions for unseen samples.

Note that we have introduced several so-called hyperparameters, i.e., the loss function, the back-propagation algorithm and the activation function, that come with a certain freedom of choice.
In this work we employ a mean-square error (MSE) loss function, the AdamW algorithm \cite{Losh17}, and choose a rectified linear unit (ReLU) \cite{Nair10} activation function.
The latter has been found to yield a better performance compared to a sigmoid activation function.
 
\paragraph{Network Design}

When it comes to designing an ANN the first thing that needs to be addressed is the network topology and the structure of the data samples, as they define the size of the first (input) layer and the last (output) layer.
While one neuron is sufficient for the output layer, which is supposed to provide a single real number that corresponds to the prediction for the converged observable, the ground-state energy in this case, one can think of different sets of input data that can be fed to the input layer of the ANN.
In previous single-nucleus applications Negoita et al.~\cite{NeVa19} have established the two input values $\hbar\Omega$ and $N_\mathrm{max}$ and have trained their ANNs to predict or parameterize the results of NCSM calculations as a function of model-space size and HO frequency. Eventually, the trained ANNs are used to predict the result of an NCSM calculation at a large $N_\mathrm{max}$, expected to coincide the with converged result. 
The functional dependence varies with nucleus and interaction and can, therefore, not be generalized.

In this work we aim for a universal ANN tool based on a different concept. 
We predict the actual converged value of an observable from a set of NCSM calculations in truncated model spaces.
Therefore, we construct data samples from $X$ sequences of NCSM calculations for different values of $\hbar\Omega$ with $L$ data points for consecutive $N_\mathrm{max}$, resulting in a total of $X\cdot L$ input neurons.
For the prediction of the converged energy based on short sequences we choose a network topology with three hidden layers with $4 X\cdot L$ , $4 X\cdot L$ and $8 X$ neurons as depicted in Fig.~\ref{fig:ANN}.

We discuss two different input modes, i.e., types of data formatting, to networks with slightly different topologies.
First, we feed the raw energy eigenvalues into the ANN.
The network resembles a mapping $M_\mathrm{ABS}: S_\mathrm{ABS}\rightarrow E^{\infty}$ where $E^\infty$ is the converged energy and $S_\mathrm{ABS}$ is an input sample of the shape
\begin{align}
    S^{\curly{N}_\mathrm{max}}_\mathrm{ABS}=\Big(&E^{\curly{N}_\mathrm{max}-6}_{\hbar\Omega_1},E^{\curly{N}_\mathrm{max}-4}_{\hbar\Omega_1},E^{\curly{N}_\mathrm{max}-2}_{\hbar\Omega_1},E^{\curly{N}_\mathrm{max}}_{\hbar\Omega_1},\notag\\
    &E^{\curly{N}_\mathrm{max}-6}_{\hbar\Omega_2},E^{\curly{N}_\mathrm{max}-4}_{\hbar\Omega_2},E^{\curly{N}_\mathrm{max}-2}_{\hbar\Omega_2},E^{\curly{N}_\mathrm{max}}_{\hbar\Omega_2},\\
    &E^{\curly{N}_\mathrm{max}-6}_{\hbar\Omega_3},E^{\curly{N}_\mathrm{max}-4}_{\hbar\Omega_3},E^{\curly{N}_\mathrm{max}-2}_{\hbar\Omega_3},E^{\curly{N}_\mathrm{max}}_{\hbar\Omega_3}\Big)\notag
\end{align}
with $\curly{N}_\textrm{max}$ being the highest $N_\mathrm{max}$ in the sample and $E^{N_\mathrm{max}}_{\hbar\Omega}$ being the NCSM energy eigenvalue for HO frequency $\hbar\Omega$ at a given $N_\mathrm{max}$.
This mode is referred to as ``ABS''. We choose $X=3$ and $L=4$ and the hidden layers are scaled accordingly.
The colored data points in Fig.~\ref{fig:NCSM} resemble one possible input sample along with the corresponding converged energy as a target value.
Note that the parameter $L$ limits the applicability of the ANN for heavier systems, as sufficiently large values of $N_\mathrm{max}$ need to be accessible through the NCSM. 
In our setup we need data at least up to $N_\mathrm{max}=8$ in order to be able to provide a prediction, since we choose to discard the $N_\mathrm{max}=0$ results.

A potential limitation of the ABS mode is the specific energy range of the input data. 
Heavier nuclei exhibit a broad range of ground-state energies that exceeds the -2 to -30~MeV range of the few-body systems that are employed in the training. One might consider to use the ABS mode in connection to the ground-state energies per nucleon to remedy this issue and we will come back to this idea later. 

Further, we design a second mode, referred to as ``DIFF'', for which the input is constructed from the energy differences between two consecutive $N_\mathrm{max}$ points.
With this concept we aim to avoid any dependencies on the energy range in order to further improve the ANN predictions.
This mode resembles a mapping $M_\mathrm{DIFF}: S_\mathrm{DIFF}\rightarrow \Delta^{\infty}$
where the input samples are given by
\begin{align}
    S^{\curly{N}_\mathrm{max}}_\mathrm{DIFF}=\Big(&\Delta^{\curly{N}_\mathrm{max}-4}_{\hbar\Omega_1},\Delta^{\curly{N}_\mathrm{max}-2}_{\hbar\Omega_1},\Delta^{\curly{N}_\mathrm{max}}_{\hbar\Omega_1},\notag\\
    &\Delta^{\curly{N}_\mathrm{max}-4}_{\hbar\Omega_2},\Delta^{\curly{N}_\mathrm{max}-2}_{\hbar\Omega_2},\Delta^{\curly{N}_\mathrm{max}}_{\hbar\Omega_2},\\
    &\Delta^{\curly{N}_\mathrm{max}-4}_{\hbar\Omega_3},\Delta^{\curly{N}_\mathrm{max}-2}_{\hbar\Omega_3},\Delta^{\curly{N}_\mathrm{max}}_{\hbar\Omega_3}\Big)\notag
\end{align}
with $\Delta^{N_\mathrm{max}}_{\hbar\Omega} = E^{N_\mathrm{max}}_{\hbar\Omega}-E^{N_\mathrm{max}-2}_{\hbar\Omega}$.
The networks output will then be the difference from the lowest lying input energy to the predicted converged value, i.e.
\begin{align}
    \Delta^\infty=E^\infty-\min(E^{\curly{N}_\mathrm{max}}_{\hbar\Omega}) &&\mathrm{for}\;\; \hbar\Omega\;\; \mathrm{in}\;\; S^{\curly{N}_\mathrm{max}}_\mathrm{DIFF}.
\end{align}
Note that the input layer for this mode has only $X\cdot (L-1)$ input neurons to ensure that both ANN types employ the same number of raw NCSM results for each sample. 
Analogously, the hidden layers are also scaled with $(L-1)$ instead of $L$.

\paragraph{Data Preparation and Training}
As mentioned above, we aim to predict the ground-state energies of a broad range of p-shell nuclei based on data from few-body systems.
Our training data consists of \elem{H}{2}, \elem{H}{3}, and \elem{He}{4} calculations up to $N_\mathrm{max}=50$, $40$, and $24$, respectively, for seven HO frequencies from $\hbar\Omega=12$ to $32~\mathrm{MeV}$.
The calculations were carried out for a family of non-local NN+3N interactions from chiral EFT at N$^2$LO, N$^3$LO and N$^4$LO' and for three cutoffs $\Lambda=450$, 500, and 550~MeV \cite{EnMa17,HuVo20}.
All interactions have been applied bare and SRG evolved with three different flow parameters $\alpha=0.02~\mathrm{fm}^4$, $0.04~\mathrm{fm}^4$, and $0.08~\mathrm{fm}^4$.
This accumulates to a total of 756 converging sequences, on which the ANNs can be trained.
In order to study the impact of the selection of the training data, we also employ different physics-motivated filters on this data set, which are discussed in the result sections.
For using the sequences in the training process, they need to be converted into samples that match the input layer of the networks.
For each nucleus and interaction all subsets of $X$ frequencies and their permutations are generated, before samples of $L$ consecutive $N_\mathrm{max}$ are constructed. As mentioned earlier, we exclude $N_\mathrm{max}=0$ results from the training. 

\begin{figure}
    \hspace{-.4cm}
    \includegraphics[width=1.06\columnwidth]{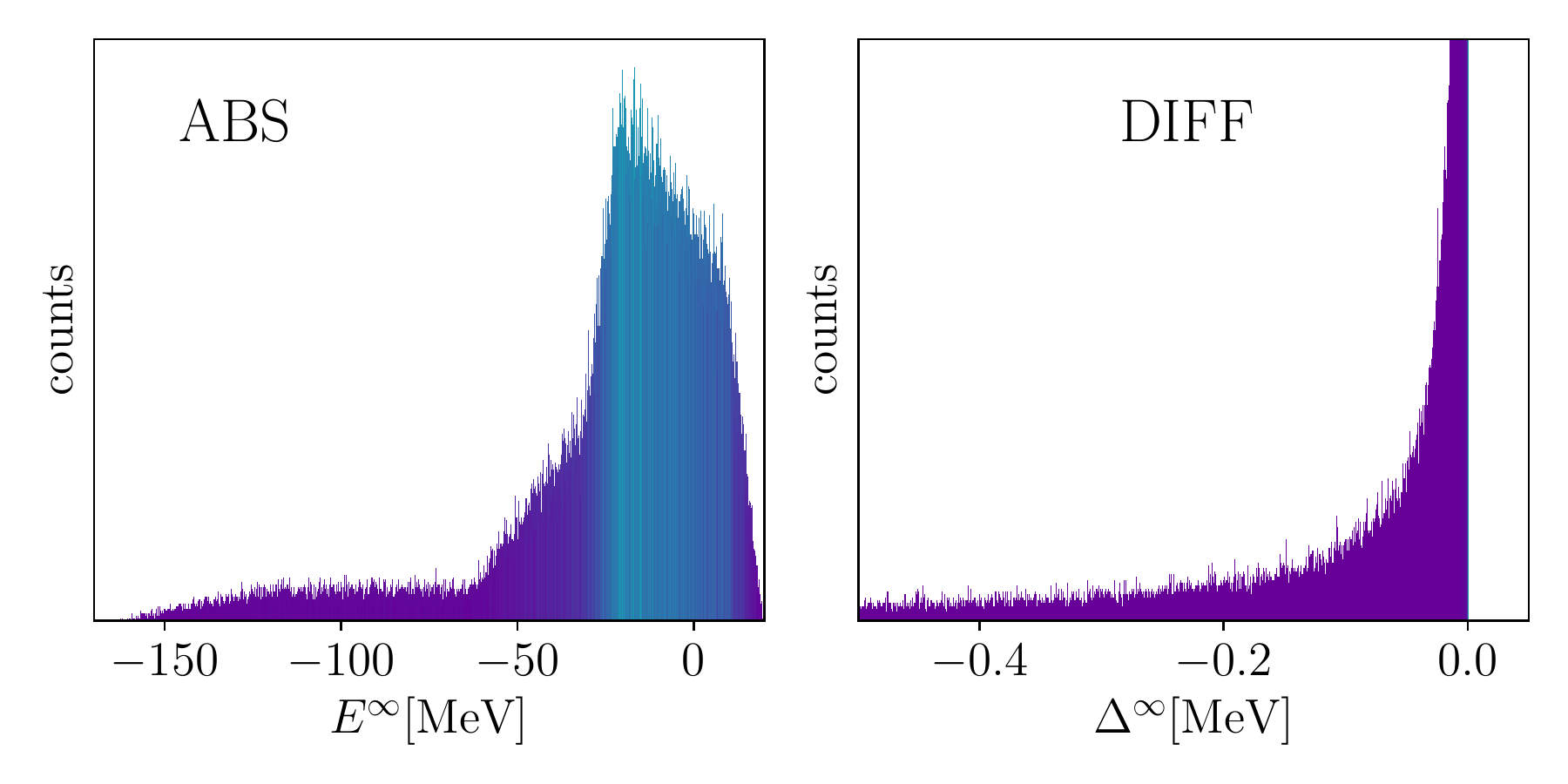}
    \caption{Distributions of the target output values of all training samples for ABS and DIFF mode. A random scaling has been applied in order to spread the values over a larger energy range to avoid training the ANN on specific values. See text for details. \label{fig:SAMPLEHIST}}
\end{figure}

Since more data is beneficial for the quality of the trained ANN, additional samples are generated by randomly scaling and shifting the existing samples via
\begin{align}
    a\cdot S^{\curly{N}_\mathrm{max}}_\mathrm{ABS}+b \quad\text{and}\quad a\cdot S^{\curly{N}_\mathrm{max}}_\mathrm{DIFF}
\end{align}
for ABS and DIFF modes, respectively, where $a\in(0.25,4)$ and ${b\in(-20,20)~\mathrm{MeV}}$ for ABS and $a\in(0.5,2)$ for DIFF are random numbers that are uniformly distributed over the given intervals.
In this way an arbitrary number of training samples, which slightly differ in their convergence behavior and are spread over a broader energy range, can be generated.
It also prevents the ANN from learning and reproducing only the three target values corresponding to the three training nuclei, which could be expected since all employed interactions produce very similar predictions for the ground-state energies of these nuclei. 
The distributions of the target values, shown in Fig. \ref{fig:SAMPLEHIST}, indicate that the majority of training samples is still in the range of the few-body systems, but also samples with a ground-state energies down to $-150~\textrm{MeV}$ are represented in the training data.
This is critical for the ABS mode as the ground-state energies of p-shell nuclei are significantly lower than energies of the training nuclei. Hence, we can avoid an inherent bias of the predictions.
Similar effects apply to the DIFF mode, where the scaled training data covers corrections beyond $-0.5~\textrm{MeV}$ instead of a few keV.

For the actual training three disjoint sets of samples are generated: (1) the training set consisting of $1\,000\,000$ samples, (2) the development set consisting of $10\,000$ samples, and (3) the validation set consisting of $50\,000$ samples.
The ANNs are trained on the training set in batches of 512 samples at a time and we iterate through the training data a total of 20 times.
During this process the average loss is monitored via an evaluation of the development set after every iteration in order to adjust any loss-dependent parameters like the learning rate, which is initialized at 0.01.
Once the training is complete, we decide to keep or retrain the network based on its performance on the validation set.
We found that the training is very robust and rarely any networks have to be discarded.
The ANNs reach an average accuracy of about 60~keV (6~keV) for the ABS (DIFF) mode on the validation set.
This indicates that the description of the training samples is more accurate when given as differences, due to the much smaller range of input values compared to the ABS method. 

\paragraph{Statistical Evaluation}
The evaluation of the ANNs is done analogously to the training.
We start from a small set of evaluation data from NCSM calculations for a specific nucleus, interaction, and eigenstate.
Again, all possible subsets of $X=3$ HO frequencies and their permutations are generated, from which input samples of $L=4$ consecutive $N_\mathrm{max}$ are constructed.
The samples are sorted by $\curly{N}_{\max}$ which describes the largest value of $N_{\max}$ present in the given sample.
A network then evaluates all samples $S^{\curly{N}_{\max}}$ for one given $\curly{N}_{\max}$ at a time.
By increasing $\curly{N}_{\max}$ we construct a sequence of predictions using information from successively larger model spaces.
In many relevant applications, the evaluation data will be limited to moderate values of $N_{\max}$, typically $N_{\max} \leq 12$ or even less. 
Therefore, we restrict our investigations to evaluation samples with $\curly{N}_{\max}\leq 12$.

\begin{figure}[b]
    \hspace{-.4cm}
    \includegraphics[width=1.06\columnwidth]{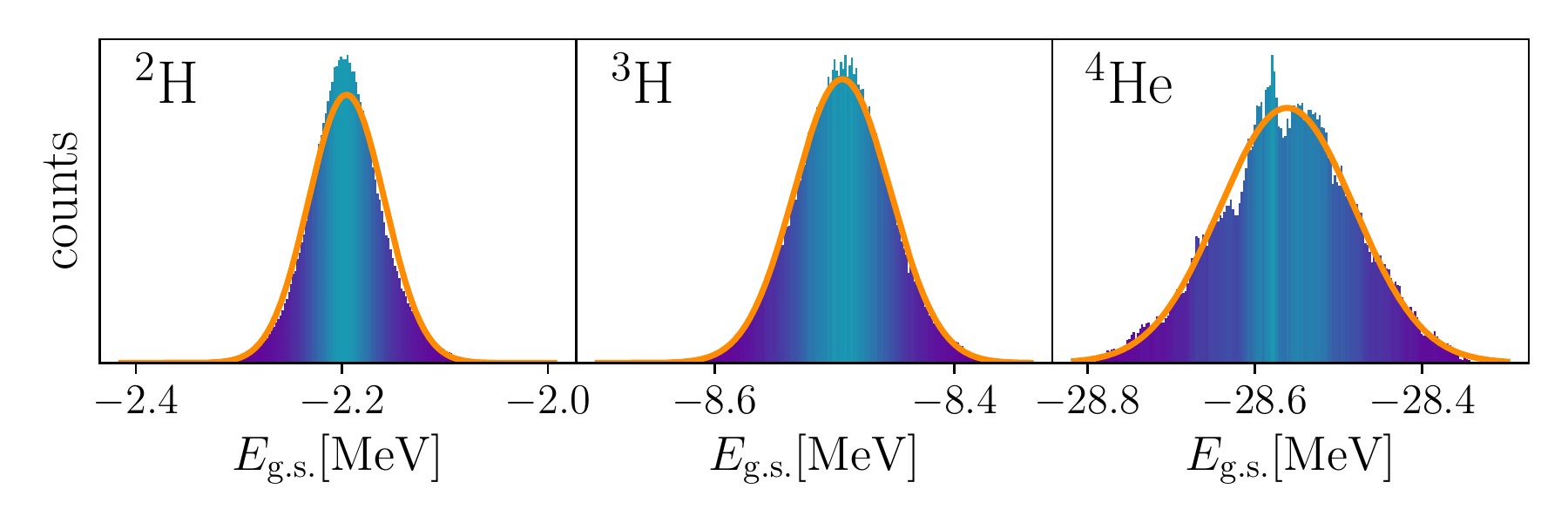}
    \caption{
    Histograms of exemplary predictions from 1000 ANNs for \elem{H}{2}, \elem{H}{3} and \elem{He}{4} at $\curly{N}_\textrm{max}=12$ with fitted Gaussian distributions.
    \label{fig:HIST}}
\end{figure}

As one finds multiple samples for a given $\curly{N}_{\max}$, we obtain multiple predictions from a single ANN, which we treat as equally plausible.
In addition, we construct a total of 1000 networks with comparable performance on the validation set and run all the evaluation samples through all networks.  
The resulting histogram of all predictions obtained in this statistical evaluation shows a single peak structure reminiscent of a Gaussian distribution. Therefore, the final prediction for the ground-state energy can be characterized through a mean value and a standard deviation---this serves as nominal prediction for the converged ground-state energy along with a quantified statistical uncertainty.
Figure~\ref{fig:HIST} shows examples for predictions for the training nuclei along with a Gaussian distribution obtained from the mean value and standard deviation, which represent the distribution well.
In case of a multi-peak structure in the histogram one should look for an imbalance in either the training data or the evaluation data as discussed in \cite{JiHa19}.

\paragraph{Application to Few-Body Systems}
First, we will take a look at the networks' performance for the few-body nuclei used in the training. For investigations of these nuclei we need to ensure that we do not reproduce any of the training samples.
We circumvent this by generating the evaluation data from another family of interactions that was not used for the training data.
All NCSM results shown in the following are obtained from a realistic semi-local NN+3N interaction from chiral EFT at N$^2$LO with cutoff $\Lambda=450$~MeV \cite{ReKr18,MeEp21} and SRG flow parameter $\alpha=0.08$~fm$^4$.

As mentioned earlier, we artificially truncate the evaluation data set at $\curly{N}_{\max} = 8,10,12$ to probe the robustness and consistency of the ANN prediction. We further employ different physics-motivated filters on the training and evaluation data in order to study the dependence of the predictions on the data selection. These are (a) all data available, (b) limitation of training data to $N_\mathrm{max}\leq12$ as this is the relevant range for actual applications, and (c) omission of training data obtained with bare interactions, i.e., interactions that are not SRG evolved, as SRG transformed Hamiltonians are used in the applications.

\begin{figure}[t]
    \hspace{-.3cm}
    \includegraphics[width=1.05\columnwidth]{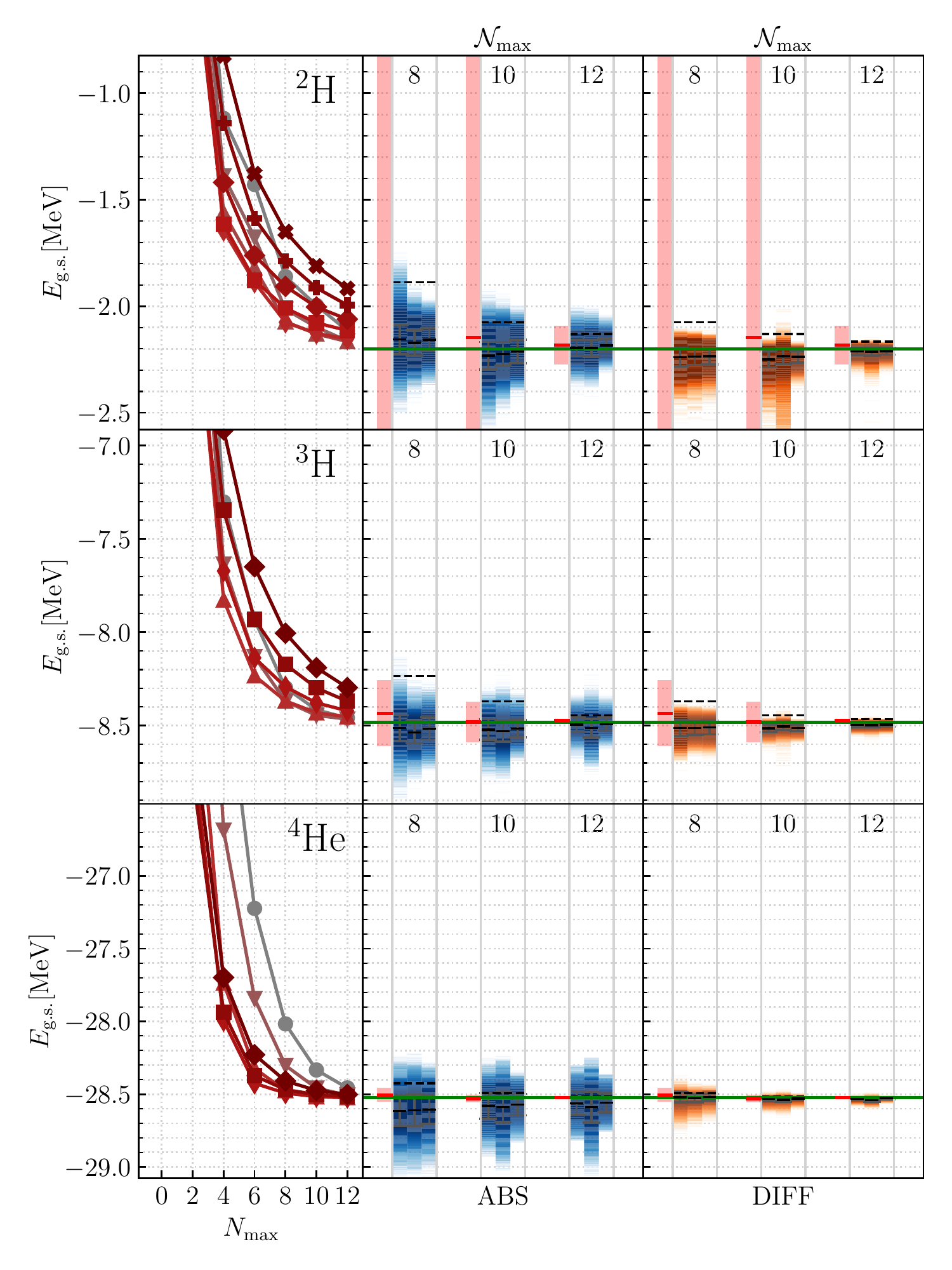}
    \caption{Input data and predicted ground-state energies for \elem{H}{2}, \elem{H}{3} and \elem{He}{4}.
    The left-hand column shows the input data for multiple HO frequencies around the variational minimum.
    In the middle and right-hand columns show the statistical predictions are shown as color-coded vertical histograms at a given $\curly{N}_\textrm{max}$.
    The groups of three histograms corresponds to the filters (a) to (c) on the training data (see text for details). The networks were trained and evaluated with either the ABS (middle) or DIFF (right) mode. The dashed black lines (\textbf{-\,-\,-}) are the variational boundaries for the respective model space and the horizontal green line \mbox{(\textcolor{FGgreen}{\textbf{---}})} indicates the converged value. For comparison classical extrapolations are given as red based with uncertainty bands (\textcolor{FGred}{\textbf{---}}).\label{fig:RES_TRAIN}}
\end{figure}

Figure~\ref{fig:RES_TRAIN} shows the results for the few-body nuclei and the corresponding numerical values are given in Tab.~\ref{tab:summary}.
The left-hand panels show the evaluation data that has been used as input for the ANNs, and the panels in the middle and on the right-hand side show the distributions of predictions from the ABS and DIFF modes, respectively.
The solid black lines with error bars indicate the mean and standard deviation of the predictions.
For comparison, classical extrapolations that are performed analogously to the procedure described in \cite{LENPIC21} are given as red bars with error bands. The green horizontal lines indicate the converged values, which have been obtained for very large $N_{\max}$.

First of all, we observe that the ANN predictions are in good agreement with the known converged ground-state energies.
Furthermore, two important features emerge: the variational boundaries are respected by the averaged predictions and the predictions become more accurate with increasing $\curly{N}_\textrm{max}$, while the uncertainty estimates are consistent with each other. Hence, we can obtain robust predictions from small model spaces already.
On closer inspection, we find that the DIFF method yields significantly smaller uncertainties and, thus, more precise results compared to the ABS method, while the predictions are equally or even more accurate with respect to the known converged energies.
One reason for the improved precision of the DIFF mode might be that the range of input values seen by the network is much smaller when using energy differences and does not change systematically from one nucleus to the other. In a broader sense this resembles a normalization of the data as it is often employed in machine learning applications \cite{Lecun12}.
 
As mentioned earlier, we can apply the ABS method to the ground-state energies per nucleon instead of the ground-state energy with the aim to reduce the range of possible target energies. We refer to this variant as ABS' and summarize results in table~\ref{tab:summary}. 
Contrary to expectations this scheme does not yield improvements compared to the ABS method, neither w.r.t. the mean predictions nor regarding the uncertainty estimates.
It even produces a significant overbinding for some nuclei in small model spaces.
Therefore, we refrain from discussing this scheme in more detail.
 
Compared to the classical extrapolations the ANN predictions, especially for the DIFF method, are more consistent while being in good agreement with the classical extrapolations.
Note that the uncertainties for the classical extrapolations for \elem{H}{2} up to $\curly{N}_{\max}=10$ are unreasonably large.
This is due to the deuteron-specific convergence behavior that cannot be modeled well with exponential functions. 

Regarding the filtered data sets (b) and (c) we find very little deviation from the unrestricted data set (a).
Hence, the ANNs seem to be able to identify the required information from the training data making them robust against changes of the training set. In other words, the predictions are not significantly biased by the choice of training data.
We note that there is no systematic improvement from a manual preselection of the best HO frequencies for the evaluation as it is done for the classical extrapolation. However, evaluating the ANNs with frequencies far from the variational energy minimum provides less accurate predictions and, therefore, the input data should use frequencies around this optimum.

\begin{figure}[t]
    \hspace{-.3cm}
    \includegraphics[width=1.05\columnwidth]{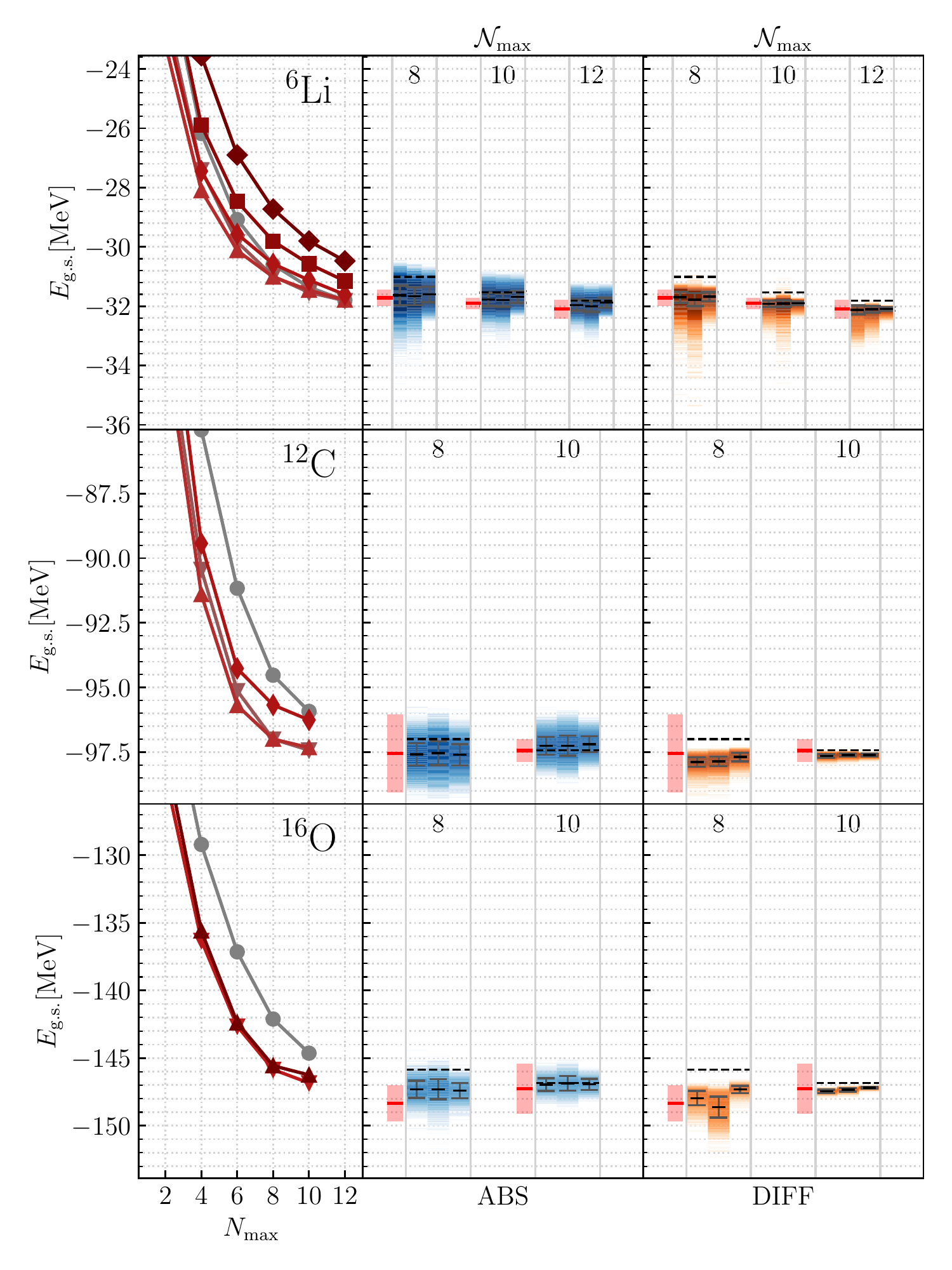}
    \caption{Same as Fig.~\ref{fig:RES_TRAIN} but for \elem{Li}{6}, \elem{C}{12} and \elem{O}{16}.\label{fig:RES_EVAL}}
\end{figure}

\hspace{-6mm}
\begin{table}[h!]
    \begin{tabular}{ c l l l }
     \hline\hline
      & \multicolumn{1}{c}{$\curly{N}_\mathrm{max}=8$} & \multicolumn{1}{c}{10} & \multicolumn{1}{c}{12} \\
     \hline\\[-1em]
     \multicolumn{4}{c}{\elem{H}{2}} \\
     \hline
     \multicolumn{1}{c|}{var. bound.} & -1.887 & -2.075 & -2.130 \\
     \multicolumn{1}{c|}{extrapol.}   & -2.720(*) & -2.147(*) & -2.183(91) \\
     \multicolumn{1}{c|}{ABS}         & -2.156(69) & -2.223(64) & -2.196(37) \\
     \multicolumn{1}{c|}{ABS'}        & -2.144(67) & -2.217(60) & -2.191(28) \\
     \multicolumn{1}{c|}{DIFF}        & -2.239(43) & -2.250(38) & -2.212(17) \\
     \multicolumn{1}{c|}{conv. result} & -2.200 & -2.200 & -2.200 \\
     \hline\\[-1em]
     \multicolumn{4}{c}{\elem{H}{3}} \\
     \hline
     \multicolumn{1}{c|}{var. bound.} & -8.233 & -8.369 & -8.445 \\
     \multicolumn{1}{c|}{extrapol.}   & -8.434(176) & -8.480(109) & -8.473(10) \\
     \multicolumn{1}{c|}{ABS}         & -8.513(68) & -8.521(55) & -8.494(40) \\
     \multicolumn{1}{c|}{ABS'}        & -8.503(76) & -8.516(47) & -8.495(29) \\
     \multicolumn{1}{c|}{DIFF}        & -8.514(42) & -8.515(19) & -8.497(13) \\
     \multicolumn{1}{c|}{conv. result} & -8.481 & -8.481 & -8.481 \\
     \hline\\[-1em]
     \multicolumn{4}{c}{\elem{He}{4}} \\
     \hline
     \multicolumn{1}{c|}{var. bound.} & -28.425 & -28.493 & -28.517 \\
     \multicolumn{1}{c|}{extrapol.}   & -28.505(48) & -28.530(24) & -28.525(9) \\
     \multicolumn{1}{c|}{ABS}         & -28.615(103) & -28.581(87) & -28.562(81) \\
     \multicolumn{1}{c|}{ABS'}        & -28.556(124) & -28.580(80) & -28.567(67) \\
     \multicolumn{1}{c|}{DIFF}        & -28.521(36) & -28.533(15) & -28.532(8) \\
     \multicolumn{1}{c|}{conv. result} & -28.524 & -28.524 & -28.524 \\
     \hline\\[-1em]
     \multicolumn{4}{c}{\elem{Li}{6}} \\
     \hline
     \multicolumn{1}{c|}{var. bound.} & -31.014 & -31.535 &  -31.813\\
     \multicolumn{1}{c|}{extrapol.}   & -31.718(278) & -31.906(189) & -32.098(317) \\
     \multicolumn{1}{c|}{ABS}         & -31.653(324) & -31.770(227) & -31.964(185) \\
     \multicolumn{1}{c|}{ABS'}        & -31.804(437) & -31.915(250) & -32.134(165) \\
     \multicolumn{1}{c|}{DIFF}        & -31.699(186) & -31.919(106) & -32.126(147) \\
     \hline\\[-1em]
     \multicolumn{4}{c}{\elem{C}{12}} \\
     \hline
     \multicolumn{1}{c|}{var. bound.} & -97.00 & -97.42 & \multicolumn{1}{c}{--} \\
     \multicolumn{1}{c|}{extrapol.}   & -97.55(150) & -97.43(44) & \multicolumn{1}{c}{--} \\
     \multicolumn{1}{c|}{ABS}         & -97.59(43) & -97.25(35) & \multicolumn{1}{c}{--} \\
     \multicolumn{1}{c|}{ABS'}        & -98.54(48) & -97.59(35) & \multicolumn{1}{c}{--} \\
     \multicolumn{1}{c|}{DIFF}        & -97.88(18) & -97.64(7) & \multicolumn{1}{c}{--} \\
     \hline\\[-1em]
     \multicolumn{4}{c}{\elem{O}{16}} \\
     \hline
     \multicolumn{1}{c|}{var. bound.} & -145.85 & -146.84 & \multicolumn{1}{c}{--} \\
     \multicolumn{1}{c|}{extrapol.}   & -148.35(134) & -147.28(1.86) & \multicolumn{1}{c}{--} \\
     \multicolumn{1}{c|}{ABS}         & -147.31(64) & -146.96(47) & \multicolumn{1}{c}{--} \\
     \multicolumn{1}{c|}{ABS'}        & -149.37(78) & -147.77(45) & \multicolumn{1}{c}{--} \\
     \multicolumn{1}{c|}{DIFF}        & -147.96(52) & -147.46(12) & \multicolumn{1}{c}{--} \\
     \hline\hline
    \end{tabular}
    \caption{ANN predictions for ABS and DIFF modes for all discussed nuclei. The ABS' results are obtained using the ABS mode with the ground-state energy per nucleon for training and evaluation. Variational bounds, classical extrapolations and, where available, converged results are given for comparison. Uncertainties denoted as (*) are unreasonably large due to a breakdown of the classical extrapolation method.}
     \label{tab:summary}
   \end{table}

\paragraph{Application to p-Shell Nuclei}
We now proceed to the heavier p-shell nuclei \elem{Li}{6}, \elem{C}{12} and \elem{O}{16}, which were not part of the training process. The results for ground-state energies are depicted in Fig.~\ref{fig:RES_EVAL} and the numerical values are given in Tab.~\ref{tab:summary}.
Starting with \elem{Li}{6} we still find consistent predictions that are in very good agreement with the classical extrapolations. Both, the ABS and DIFF modes produce very similar results.
We note that some of the distributions exhibit tails towards lower energies, which indicates that a Gaussian is not the ideal choice for modeling the uncertainties. One could turn to a Bayesian treatment with more complicated probability density functions as suggested in \cite{JiHa19}.

The last nuclei we discuss are \elem{C}{12} and \elem{O}{16}.
Since they are much heavier, it is more challenging to converge the NCSM calculations and, therefore, these nuclei resemble perfect examples for the applications this machine learning tool is aiming for.
We again find very consistent predictions for the ground-state energies that agree well with the classical extrapolations. 
However, some predictions for the ABS method at $\curly{N}_\textrm{max}=10$ violate the variational bounds. 
The DIFF method on the other hand produces very accurate predictions for both nuclei that are compatible with independent results presented in \cite{LENPIC21}.

For all p-shell nuclei, the predictions with the different filtered data sets do not differ much from each other. There might be a slight advantage in precision and consistency for filter (c), which omits the bare interaction results from the training, but this might change for other evaluation sets. Generally, we recommend a full survey of all filters and all accessible $\curly{N}_\mathrm{max}$ for future applications, in order to check for consistency and robustness of the predictions.

\paragraph{Conclusions}
We have developed a new machine learning tool for the prediction of converged ground-state energies based on input data from non-converged  NCSM calculations. The key idea is to construct a universal ANN to identify the convergence pattern on NCSM calculations for arbitrary nuclei and interactions and to predict the converged ground-state energy directly. The training is based on a huge set of NCSM results for few-body systems with different chiral NN+3N interactions and SRG evolutions, where in each case the converged results is precisely known.  
The evaluation of the ANNs is embedded into a statistical framework that provides predictions with quantified uncertainties. We have shown that NCSM calculations for light systems up to $A=4$ contain enough information on the convergence patterns to provide reliable predictions with competitive uncertainties for heavier nuclei. The quality of the predictions exceeds the capabilities of traditional extrapolations and is comparable with previous ANN extrapolations \cite{NeVa19}, which were trained specifically for a single nucleus and interaction. It is important to note that the ANNs presented here are only trained once, and are then ready for a broad range of applications in future calculations.

The scheme presented here can be easily transferred to other ab initio many-body approaches. It can also be generalized to other observables---work for the prediction of converged rms-radii is already in progress. In these cases the convergence behaviour is not constrained by the variational principle and, therefore, more complex. An important next development step to tackle other observables and to further improve the precision of energy predictions is, e.g., the normalization of the input sequences beyond the DIFF mode presented here. Furthermore, we can enrich the variety of training data by considering synthetic nuclei obtained with modified Hamiltonians, which cover a controlled range for converged observables. Work along these lines is in progress.      

\paragraph{Acknowledgments}

This work is supported by the Deutsche\linebreak Forschungsgemeinschaft (DFG, German Research Foundation) through the DFG Sonderforschungsbereich SFB 1245 (Project ID 279384907) and the BMBF through  
Verbundprojekt 05P2021 (ErUM-FSP T07, Contract No. 05P21RDFNB).
Numerical calculations have been performed on the LICHTENBERG II cluster at the computing center of the TU Darmstadt.

\bibliographystyle{elsarticle-num}
\bibliography{bib_nucl}

\end{document}